\documentclass[12pt,preprint]{aastex}

\shorttitle{Low-Mass Close Binaries}
\shortauthors{Basri \& Reiners}

\begin{document}


\title{A Survey for Spectroscopic Binaries Among Very Low-Mass Stars}


\author{ Gibor Basri and Ansgar Reiners\altaffilmark{1,^\star}}
\affil{Astronomy Department, University of California, Berkeley, CA 94720}
\email{[basri,areiners]@astron.berkeley.edu}
\altaffiltext{1}{Hamburger Sternwarte, Universit\"at Hamburg, Gojenbergsweg 112, D-21029 Hamburg, Germany}
\altaffiltext{$^\star$}{Marie Curie Outgoing International Fellow}




\begin{abstract}
We report on the results of a survey for radial velocity variability in a 
heterogeneous sample of very low-mass stars and brown dwarfs. One
distinguishing characteristic of the survey is its timespan, which allows
an overlap between spectroscopic binaries and those which can be found
by high angular-resolution imaging. Despite our relatively low velocity
precision, we are able to place a new constraint on the total binary
fraction in these objects, which suggests that they are more likely the
result of extending the same processes at work at higher masses into this
mass range, rather than a distinct mode of formation. Our basic result is 
that there are $6 \pm 2$ out of 53, or $11^{+0.07}_{-0.04}$\% spectroscopic
binaries in the separation range 0-6\,AU, nearly as many as resolved binaries.
This leads to an estimate of an upper limit of $26 \pm 10$\% for the binary 
fraction of VLM objects (it is an upper limit because of the possible overlap 
between the spectroscopic and resolved populations). A reasonable estimate
for the very low-mass binary fraction is $20 - 25$\%.  

We consider several possible separation and frequency distributions, 
including the same one as found for GK stars, a compressed version of that, 
a version of the compressed distribution truncated at 15 AU, and a 
theoretical distribution which considers the evaporation of small-N clusters. 
We conclude that the latter two bracket the observations,
which may mean that these systems form with intrinsically smaller 
separations due to their smaller mass, and then are truncated due to their
smaller binding energy. We do not find support for the ``ejection 
hypothesis'' as their dominant mode of formation, particularly in view of
the similarity in the total binary fraction compared with slightly more
massive stars, and the difficulty this mechanism has in producing numerous
binary systems. Our conclusions must be viewed as tentative until studies 
with larger and better-posed samples, and higher velocity precision are
conducted.

\end{abstract}





\newpage

\section{Introduction}

It has been about 15 years since the study of very low-mass objects
(VLMS) with high spectral resolution became possible, and only a
decade since the discovery of brown dwarfs \citep{Bas00}.  Henceforth
we shall not distinguish brown dwarfs separately; they are essentially
the same as very low-mass stars for the purpose of this paper, and our
sample is primarily stellar in any case.  There is no reason for the
star-formation process to know about the later ability of the forming
object to reach the main sequence, but it is true that the formation
of VLMS occurs at a scale that is much smaller than the typical Jeans
mass in a molecular cloud (which is more like a solar mass).
Nonetheless, the observed numbers of VLMS are comparable to those of
higher-mass stars \citep{Chab03}.

In light of this, it is reasonable to ask whether the formation of
VLMS utilizes the same process of collapse of dense molecular cores as
do solar-type stars.  In that case, the mass spectrum of the final
objects could follow a modified mass spectrum for cores.  Turbulent
fragmentation in molecular clouds \citep{Pad04} is one way in which
this could be achieved.  Supersonic turbulence generates a chaotic
complex velocity field, within which shocks cause the formation of
dense filaments.  Along these there are places where the local density
is high enough for gravity to overwhelm pressure support (turbulent,
thermal, and magnetic) and cause collapse to a compact object.  The
mass of the object is then set primarily by the mass available in the
unstable condensation.  In their calculation \citet{Pad04} find that
they can approximately reproduce the IMF, including the ratio of VLMS
to higher mass stars.

We have observed many of instances of very young VLMS in star-forming
regions \citep[eg.][]{JMB03, Luh04, Nat04}.  These allow us to
empirically address questions of their formation and how it compares
to that for higher mass stars.  The basic conclusion from these
studies and many others like them is that very young VLMS exhibit all
of the same phenomena that are seen in higher-mass T Tauri stars,
including similar rotations, magnetospheric accretion, and outflow
indicators.  The fractional numbers and lifetimes of accretion disks
also appear similar, but the accretion rates scale down with some
power-law index of mass \citep[perhaps $-2$;][]{MJB05}.  All this
seems to suggest that VLMS form in essentially the same fashion as the
rest of stars.

Alternatively, it has been suggested that the predominant channel of
formation for VLMS begins like that for stars, but in the context of
small-N clusters of objects.  In that case, the objects can interact
gravitationally with each other, and the lightest objects tend to be
ejected \citep{Lar72, RC01}.  If this happens before they finish their
main accretion phase, the mass of the object could be prematurely
truncated, leading to a VLMS.  This mechanism supposes, therefore,
that there would be many fewer VLMS if star-formation were largely an
isolated phenomenon.  That basic picture has received support from
calculations such as by \cite{Bate02}.  A high fraction of pre-VLMS
are produced in their computation, generally by early ejection from
small-N clusters aided by objects falling towards formation centers
along filaments.  The extent to which this happens depends on the
initial turbulent spectrum in the cloud (fragment), and there are
questions as to whether the right initial and computational conditions
have been used.  Nonetheless, they show the power of the basic
concept.

In practice this mechanism must operate at some level, so the real
question is how often it occurs compared with relatively undisturbed
stellar formation.  Put another way, would the fraction of VLMS be
substantially different without premature truncation of mass accretion
during formation by dynamical interactions?  It is not clear to what
extent the conclusions from observations of forming VLMS are in
conflict with the ejection hypothesis.  One might expect inner
accretion disks to survive ejection, but their lifetimes and
radial extents would be shortened.  Radial extents of disks have not 
yet been tested (a problem for sub-mm and mm observatories), and the 
lifetime indicators do not support early termination \cite{Jay03}. 
Tests based on ejection velocities are not conclusive (even though 
there is no evidence for generally higher velocities for VLMS).  
Given that there is probably a mix of mechanisms, any workable test 
must be fairly robust.

Many stars form in binary or multiple systems, although the rules
governing formation of multiple stars are much less well known than
those for single stars.  It is natural to ask whether the same rules
apply throughout the stellar mass range, and into the substellar
domain.  Indeed, binary formation can serve as a probe of the basic
mechanism of formation, because this is one arena in which the standard
and ejection formation hypotheses make sufficiently different
predictions as to be testable.  We know what the binary frequency and
separation distributions look like for low-mass and solar-type stars.
Thus, if the VLMS appear to continue these relations, that is in favor
of the standard formation model, while if there is a substantial
difference for VLMS stars it suggests a different mechanism.

The ejection hypothesis has two effects on binaries.  In the first
place, it is more difficult to eject binary systems whole.  Low-mass
binaries are more loosely bound at a given separation.  Even if a
binary has two low-mass components and is fairly hard (meaning
strongly gravitationally bound), interaction with a more massive
object often leads to a sort of ``charge-exchange'' interaction which
will still tend to eject the single lightest object.  Thus, one
prediction of the ejection hypothesis is a low binary fraction for
objects predominantly formed in this manner.  In the second place,
systems that are not hard enough will tend to be disrupted, either
during ejection or later by interactions with nearby objects.  This
leaves an observational signature in the form of a dearth of wider
binaries at lower masses.

Such a dearth of wide binaries is in fact observed
\citep[eg.][]{Bouy03, Close03}.  This is a very strong result, since
wider binaries would be easier to see (down to lower mass ratios as
well).  This weighs in favor of the ejection hypothesis.  The problem
here is that it could also be produced in dense star-forming regions
(where most stars are formed) by close stellar encounters.  It could
also be produced by dynamical interactions in small-N clusters which
continue after the main accretion phase (in which case the mass
spectrum would not be due to ejection, even though the binary
distribution showed its effects).  The dynamical evolution of low-mass
multiple systems has been considered in detail by \cite{SD03}, who
find that the separation truncation is a natural result.  These
scenarios also predict a tendency towards more equal mass ratios
(which is observed). This is also partly due to the increasingly
restricted range of masses available to the secondaries. It would not
be surprising if binary separations scale naturally with mass in the
standard scenario; theoretical support for this is provided by eg.
\citet{Fish04}.  There is some evidence for a smaller mean separation
for M stars \citep{FM92, Marchal03} compared with G stars
\citep{DM91}.

We therefore return to the basic binary frequency as currently the
best means of testing the two formation scenarios.  None of the
ejection theories have yet come up with a way to get the binary
frequency above something like 5\%.  Early indications that the VLMS
binary frequency might be high \citep{MB00} have been borne out for
resolved binaries \citep{Bouy03, Close03}.  These systems, which can
be resolved by HST or AO observations, sample the binary frequency
from a few AU outward, and already show a frequency of at least 15\%.
This is often quoted as the binary frequency for VLMS, but of course
it does not include the closer binaries, which might be found by
radial velocity variability.  We don't know how close binaries compare
with wider ones -- does the binary distribution look like that for G
stars \citep{DM91} squeezed to smaller separations, or does it look
like that for G stars, but truncated at 15 AU (and possibly below a
couple of AU as well), or altogether different?  In the first case the
overall binary frequency for VLMS could approach 55\%; in the second
case it might be between 20-40\%.  Almost any result presents a
substantial problem for the ejection hypothesis as the {\it dominant}
formation mode for VLMS.

Not much is known about the frequency of VLMS spectroscopic binaries
because the required medium- to high-resolution spectroscopy has not
been available until somewhat recently, and large amounts of large
telescope time have not been devoted to this problem (the targets are
all faint, and several epochs of observation are desired for best
results).  It has been clear that such binary systems do exist; the
first brown dwarf binary found (in a sample of 7 Pleiades VLMS) is a
double-lined spectroscopic binary (SB2) with a 6-day orbital period
\citep{BM99}.  Two SB2 systems were identified in a sample of 39 field
objects by \cite{Reid02} and 1/16 in Upper Sco by \citet[][it is worth
noting here in advance of publication that our similar-sized samples
in Taurus and IC 348 turned up no further obvious SB2]{RBM05}.
Surveys for radial velocity variability due to single-lined
spectroscopic binaries (SB1) have turned up a few more: 3/24 field
objects in \citet[][two of them are SB2]{GW03}, 2/11 in ChaI by
\cite{J05}, and 3/60 in $\sigma$ Ori by \cite{Ken05}. It is apparent
that a few percent of VLMS will be SB1 or SB2, so that a survey
undertaken to gain insight on the binary frequency will have some
success. One caveat is that searches for SB1 in star-forming regions
(especially those relying on single-epoch comparisons with mean
cluster velocities) suffer from the possibility that substantial
radial velocity variability may be induced by large starspots on young
objects \citep{Saar97}.

Many of the first epoch spectra we used for this project have been
taken in a number of studies of VLMS using high spectral resolution at
the Keck telescope.  We therefore undertook a program to re-observe
all the field VLMS for at least one more epoch (and further epochs
when possible, or when triggered by apparent variability after 2
epochs).  Because the original observations were not conducted
specifically to maximize radial velocity precision, we conducted the
subsequent observations with the same technique in the same settings.
This means we did not achieve the maximum precision possible, but were
able to cover all the targets in a consistent fashion.  One advantage
(in retrospect) of this program was that not much time was granted in
any given observing period, so the time coverage on the systems was
longer than for other programs.  This helps when searching for
binaries out to the orbital periods that are picked up by HST or AO
surveys.  We did not observe the actual samples used in imaging
surveys, but the overall binary statistics can be compared.  We
describe our observational sample and techniques in
\S\ref{sec:observ}, obtain the results in \S\ref{sec:results}, and
discuss our conclusions in \S\ref{sec:discuss}.

\section{Observational Sample and Techniques}
\label{sec:observ}

Our sample was largely determined by the observations discussed in
\cite{MB03}.  All observations were conducted with the HIRES echelle
on the Keck I telescope.  The sample consists of a collection of stars
M5 and later, which was assembled during the years of discovery of
VLMS, and so is rather heterogeneous.  They are all field objects, and
tend to be at the brighter end of objects discovered.  Eventually the
pace of discovery greatly outstripped our ability to observe new
objects, so the actual sample is not a well-defined subset of all
possible targets.  A few objects were also added during the course of
this program to fill in gaps in RA when possible, which do not appear
in \cite{MB03}.  The ages of the objects are widely dispersed,
although the L dwarfs are likely to be younger than about 1 Gyr.  Some
age analysis of our sample appears in \cite{MB03}.

We did not exclude objects because of rapid rotation, although it
reduces our potential precision in some cases.  A more serious concern
is the quite variable S/N present in the spectra.  The original epochs
were taken with somewhat varied purpose, and exposures were often not
as long as would be optimal.  We therefore had to discard a reasonable
fraction of the original sample after a precision determination had
been made and they were found to be below our desired threshold.
Since the lowest precision observation determines our ability to find
variability, our second epoch observations were done in similar
fashion as the first \citep[for a description of the observational
parameters, cf.][]{MB03}. A few different echelle setups had been
employed (the HIRES echelle has incomplete coverage in the red, so one
has to decide what spectral lines are crucial for a given science
objective).  We retained the settings of the first epoch for
subsequent epochs on each target.  The final sample of systems tested
for variable radial velocity contains 53 VLMS.  This constitutes the
largest sample tested in an observationally consistent way to date, 
and also samples the greatest time differentials.

We took ThAr spectra at the beginning and end of each night to monitor
the absolute wavelength scale, but not bracketing each spectrum (since
that had not been done at the first epoch).  In retrospect it might
have been worth the time to take extensive lamp spectra, but we were
aware that telluric absorption and emission lines were also present in
each spectrum, which we hoped would provide a sufficient reference
standard.  In the end, we rely only on the airglow emission lines
(using the ThAr only to determine the dispersion).  We never hoped to
achieve high radial velocity precision; one does not really need to do
better than 1 \,km\,s$^{-1}$ in order to meet the fundamental goals of
this project.

\subsection{Method of analysis}
\label{sec:analysis}

Radial velocity measurements currently reach a precision of a few
\,m\,s$^{-1}$ in bright F- and G-type stars using an Iodine cell as a
reference \citep{But96}.  That method is not applicable in our sample,
because most of our targets are very red, and fainter than 14\,mag in
$I$.  Iodine does not have useable lines in the near infrared
wavelength region we use, and the targets are even fainter in the
region where Iodine lines are present.  The great advantage of the
Iodine method, however, is that target spectrum and reference spectrum
are taken in the same exposure through the same optical path, leading
to highest accuracy in measuring velocity shift.  Another reasonably
stable reference always visible in observations of faint targets is
emission of the night sky.  The airglow spectrum is superposed onto
the target spectrum, and can be extracted from the regions on the CCD
perpendicular to the direction of dispersion.  It does not, however,
account for positioning errors or seeing variations of the star within
the slit, which have to be comparably large due to the faintness of the
targets.

To obtain a differential radial velocity, we first deduce the shifts
between the night sky spectra and the object spectra (night sky and
object spectra are contained in the same exposures) by individually
computing both correlation functions in each order. The highest
accuracy is reached using only orders containing enough lines both in
the target spectrum and the reference night sky spectrum. In the
available spectral region we identified four orders where the flux is
high enough in both the stellar spectrum and the airglow emission
lines. Regions dominated by telluric lines are masked out for this
purpose. The wavelength regions of the four orders used are given in
Table\,\ref{tab:orders}.  We calculate the shift between the night sky
spectra and the shift between the star spectra in each order. Then we
take the median of the night sky shifts from different orders and the
median of the stellar shifts. The difference of these medians is the
shift between night sky spectra and object spectra. Taking the median
of the differences instead does not affect our final result.

The differential radial velocity between two exposures is then this
shift, corrected for the barycentric velocities on the two nights of
observation.  The measurements are given in Tables\,\ref{tab:variable}
and \ref{tab:constant} in column four. The errors listed are the
standard deviation of the individual differential radial velocities
for the four orders, compared to the median for a given pair of
observations. When there are more than two observations, we sum the
errors in quadrature for all possible pairs to yield the error listed,
and the differential radial velocity listed is the maximum found among
all pairs. Checks that all differential radial velocities are
consistent with each other were performed. Individual values for all
sets of exposure are given in Table\,\ref{tab:appendix} in the
Appendix.

As a final check we also computed correlation functions using
overlapping chunks of 150 pixels each.  From more than 60 correlation
functions we checked our adopted error on scattering among those, we
always found them to be consistent.  Differential radial velocity
variations between night sky emission and the target spectra were
found to be as accurate as 300\,m\,s$^{-1}$ using this method.
Overall, however, including all errors the typical uncertainty of a
measurement was 1.3\,km\,s$^{-1}$. In two stars, 2MASS\,1506 and
2MASS\,1615, we could only use results from one order and have no
uncertainty from a scatter between different orders. We therefore
assigned our typical uncertainty to those two measurements.

\begin{deluxetable}{cc}
  \tablecaption{Orders used}
  \tablewidth{0pt}
  \tablehead{\colhead{Order number} & \colhead{approximate wavelength range}}
  \startdata
  45 & 7865 -- 7975 \AA \\
  44 & 8040 -- 8155 \AA \\
  42 & 8425 -- 8545 \AA \\
  41 & 8630 -- 8750 \AA \\
  \enddata
  \label{tab:orders}
\end{deluxetable}

To minimize light loss, our observations were carried out with a slit
size of 1.15\,arcsec ($R = 31\,000$).  The typical seeing of the order
of 0.8\,arcsec at Mauna Kea implies that the star may not be always
exactly centered in the slit, and the geometrically shifted spectrum
can mimic a shift in radial velocity.  The maximum amount of the shift
due to imperfect centering depends on the seeing and becomes worse
with better seeing (if the star always overfilled the slit, we
wouldn't have this problem).  For the maximum possible error between
two exposures at a seeing of 0.8\,arcsec, we calculate a corresponding
maximum shift in radial velocity of about 2\,km\,s$^{-1}$, i.e., if
the star was positioned at the right edge of the slit during the first
exposure and on the left edge during the second exposure, the
``virtual'' shift between the two spectra in dispersion direction on
the CCD would correspond to a radial velocity difference of
2\,km\,s$^{-1}$.

We calculated the probability distribution of these virtual radial
velocity shifts introduced by imperfect centering of the target.
Assuming a random distribution of the star's position in the slit, we
find that at a typical seeing of 0.8\,arcsec the shift is smaller than
1.3\,km\,s$^{-1}$ in 70\,\% of all observations (note that the match
with the 1.3\,km\,s$^{-1}$ used above as intrinsic accuracy for the
stars with only one measured order is purely incidental). We expect
this effect to be effectively smaller in our observations since
spectra of faint objects are smeared out in the dispersion
direction due to imperfect tracking, i.e. our effective seeing is
worse than the nominal seeing at Mauna Kea.

An independent estimate of the radial velocity shift introduced by
imperfect centering can be calculated from the shift of telluric
absorption lines.  These lines are quite stable at the level of a few
hundred \,m\,s$^{-1}$.  The most useful region containing telluric
absorption lines is the oxygen A-Band around $\lambda\,7630$\,\AA.  In
spectra that contain this region we also calculated the shift between
the telluric lines in two spectra.  Their difference gives us an
estimate of the shift introduced between two spectra due to an
imperfect centering of the star. The reason why we do not employ these
absorption lines as a radial velocity reference is that they do not
have broad coverage in our spectra and we cover different regions in
different setups.  A calculation of differential velocity shifts
within each order is much more accurate than inter-order comparisons.
In any case, in our faint object spectra airglow lines are better
defined than the telluric lines and less subject to night to night and
airmass variations. In many objects we were not able to calculate 
shifts from telluric lines at all; sometimes our spectra simply do 
not cover useful regions of telluric absorption lines.

We calculated a mean absolute virtual shift (due to the imperfect
centering of the star in the slit) from the stars in which the
reference value of the shift has been derived from telluric bands.
Individual values for each star are given in column five of
Tables\,\ref{tab:variable} and \ref{tab:constant}. The mean absolute
telluric shift is 1.0\,km\,s$^{-1}$, and the median absolute telluric 
shift is 0.8\,km\,s$^{-1}$. As expected, these values are smaller than 
the 1.3\,km\,s$^{-1}$ mentioned above, confirming our intuition about 
the positioning uncertainty. It turns out that the estimate of the binary
frequency is sensitive to which value we choose (as explained below). We 
therefore take 0.9\,km\,s$^{-1}$ as our adopted error in telluric shifts,
and consider that this is itself uncertain by 0.1\,km\,s$^{-1}$. 

For our final radial velocity variation, $\Delta v_{\rm total}$, we
take the difference between the reference shift from telluric lines,
$\Delta v_{\rm tell}$ and the measured radial velocity variation of
the target, $\Delta v_{\rm star}$ (using zero shift correction in
cases without a telluric measurement). The final error on this shift
is found by adding the internal error (scatter from the 4 orders) to
the estimated systematic telluric error in quadrature. The final
radial velocity variation, $\Delta v_{\rm total}$, is given in column
six of Tables\,\ref{tab:variable} and \ref{tab:constant}.

\subsection{Significance of radial velocity variations}

In the eighth column of Tables\,\ref{tab:variable} and
\ref{tab:constant}, we list for each object the probability $p$ that a
random measurement of a radial velocity shift $\Delta v_{\rm total}$
in a single star (i.e., with constant radial velocity) is smaller than
the one we have measured. We computed $p$ from the uncertainty (the
updated uncertainty in cases without a telluric reference shift) in
$\Delta v_{\rm total}$ and assuming a normal distribution of
measurement errors.

The significance of a measured radial velocity variation is taken from
the probability $p$. Since we are assuming a normal distribution of
measurement errors, $p$ is assumed to underlie a normal distribution
with one degree of freedom. We classified a detection of radial
velocity variability in a star as significant if the measured radial
velocity shift was larger than 95.4\% ($2\,\sigma$) of what would be
observed in random measurements assuming no radial velocity shift and
a normal distribution with the uncertainty given for that object. This 
is a less conservative standard than often is applied, and it is possible
that 2 of the ``significant'' variable radial velocities are spurious.
As can be seen below, however, the situation is not that simple.

The assumed error on the telluric shift turns out to be crucial. If we
use the mean value of 1.0\,km\,s$^{-1}$, then 4 targets lie above the
2 sigma significance limit. If instead we use the median value of
0.8\,km\,s$^{-1}$, then 8 targets are significant variables. We
compromise on the average of these two possible errors
(0.9\,km\,s$^{-1}$), and indeed obtain the intermediate result of 6
significant radial velocity variables. At the suggestion of the referee,
we looked at the distribution $P(\chi ^2)$ for various assumed values
of the telluric error. The expected flat distribution in that quantity
occurred for a telluric error of 0.5\,km\,s$^{-1}$. We prefer not to be
that optimistic, however. The best statement about this telluric error
sensitivity seems to be that we have found $6 \pm 2$ spectroscopic
binaries in this project.

These six objects (11\%) are listed in the upper part of
Table\,\ref{tab:variable}. Since the probablilities are given in
order, one can infer the smaller or larger possible sets of variable 
stars. In the lower part of Table\,\ref{tab:variable}, we list the 
objects with measurements larger than 68.3\% of random measurements
($1\,\sigma$). Stars with smaller values of $p$ are given in
Table\,\ref{tab:constant}.
 
\begin{figure}
  \plotone{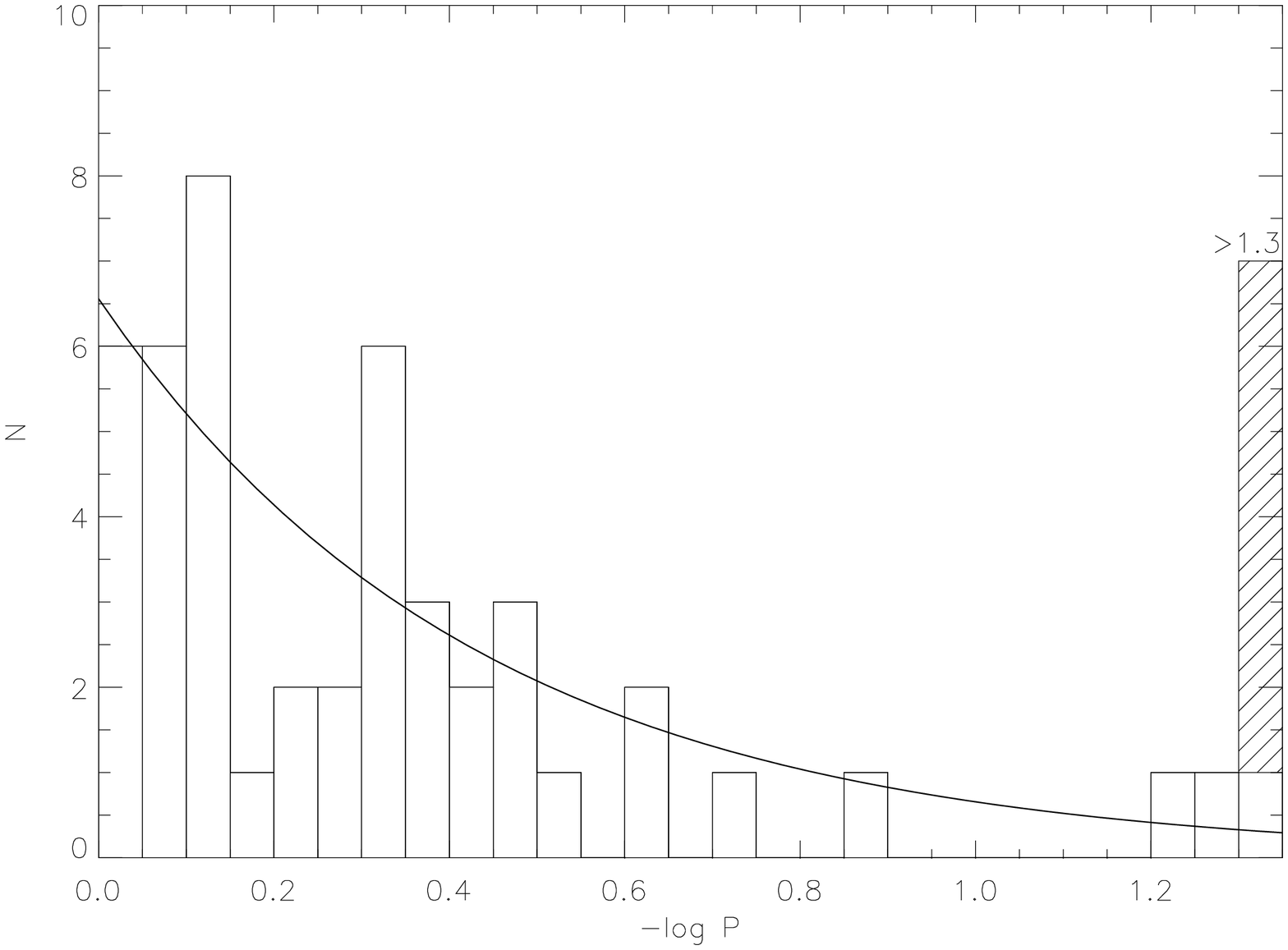} \figcaption{\label{fig:logP} Distribution of
    $\log\,P = \log\,(1-p)$ for our sample together with the expected
    distribution for non-variable stars given our uncertainties. We
    classify objects as variable if $\log\,P < -1.347$, the rightmost
    bin contains all objects with $\log\,P < -1.3$, The six objects
    classified as variable are marked with a hatched region.}
\end{figure}

We plot our distribution of probabilities $\log P = \log (1-p)$
together with the expected distribution of $\log\,P$ for our
measurement uncertainties in Fig.\,\ref{fig:logP}. All stars with
$\log\,P < -1.3$ are plotted in the last bin. The threshold for our
classification of variability is $\log\,P < -1.347$, the six objects
are marked with a hatched region. We see no obvious deviation of the
targets classified as non-variable from the expected distribution.
However, the last bin containing 7 objects is significantly
overpopulated, which is consistent with the assumption that the six
objects in the hatched region are variable. From Fig.\,\ref{fig:logP},
we see no reason to believe that a significant amount of non-variable
objects are classified as variable.

Our sample consists of 53 objects; 6 of them show significant radial
velocity variations which we ascribe to their being members of a
binary system. Although this is the largest sample of radial
velocity surveys in very low-mass objects so far, the absolute number
of probed targets (and detections) still must be considered small
in a statistical sense. Thus, in order to estimate the uncertainty of
the measured binary fraction, we searched for the fraction of binaries
at which the binomial distribution for 6 positive events in 53
performances falls off to 1/$e$ of its maximum value, which is at the
fraction of 11\%.  As a result, we get a binary fraction with errors
in our sample of $11^{+0.07}_{-0.04}$\%. This is much the same result 
as from taking $6 \pm 2$ variables out of 53 targets as the direct
result and error estimate.

\begin{deluxetable}{rlccrccc}
  \tablecaption{\label{tab:variable} Stars with indications for radial
    velocity variabilities} \tablecolumns{8} \tablewidth{0pt}
  \tablehead{\colhead{Object} & \colhead{Spectral} & \colhead{No. of} & \colhead{$\Delta v_{\textrm {star}}$} & \colhead{$\Delta v_{\textrm {tell}}$} & \colhead{$\Delta v_{\textrm {total}}$}  & \colhead{max. $\Delta t$} & \colhead{$p$}\\
   & Type & Spectra & [km/s] & [km/s] & [km/s] & [d] &} \startdata
  \multicolumn{8}{c}{2$\sigma$ detections}\\
  \noalign{\medskip} \hline \noalign{\medskip}
  LHS~3495    &  M5.5 & 2 & $1.8  \pm  0.1 $    &   $-$0.2 & $2.0  \pm  0.9 $&    709 & 0.973\\
  LHS~2645    &  M7.5 & 4 & $3.3  \pm  0.6 $    &        & $3.3  \pm  1.5 $&   3649 & 0.975\\
  DENIS~0021  &  M9   & 2 & $3.0  \pm  0.9 $    &   $-$0.1 & $3.1  \pm  1.3 $&   1135 & 0.985\\
  BRI~0021    &  M9.5 & 5 & $5.8  \pm  2.0 $    &    1.4 & $4.4  \pm  2.2 $&   3283 & 0.955\\
  DENIS~1626  &  M    & 5 & $1.3  \pm  0.3 $    &   $-$1.0 & $2.3  \pm  1.0 $&   1445 & 0.989\\
  2MASS~1506  &  L3   & 2$\tablenotemark{b}$ & $2.4 \pm 1.3$   &  -2.4 & $4.8 \pm 1.6$&  1415  & 0.998\\
  \cutinhead{1$\sigma$ variabilities}
  YZ~CMi$\tablenotemark{a}$ &  M5   & 2 & $1.6  \pm  0.1 $ && $1.6 \pm 1.4$& 2901 & 0.764\\
  LHS~248     &  M6   & 3 & $1.3  \pm  1.3 $    &   $-$1.7 & $3.0  \pm  1.6 $&  4137 & 0.942\\
  LHS~3339    &  M6   & 2 & $0.6  \pm  0.2 $    &    2.4 & $1.8  \pm  0.9 $&    336 & 0.949\\
  DENIS~1148  &  M8   & 3 & $0.8  \pm  0.3 $    &    1.9 & $1.1  \pm  1.3 $  &  1108 & 0.754\\
  2MASS~1242  &  M8   & 2 & $0.2  \pm  0.3 $    &   -0.8 & $1.0  \pm  1.0 $  &  1459 & 0.708\\
  LHS~2924    &  M9   & 4 & $2.2  \pm  0.6 $    &        & $2.2  \pm  1.5 $&   2309 & 0.864\\
  2MASS~0746  &  L0.5 & 2 & $2.1  \pm  1.8 $    &   $-$0.5 & $2.6  \pm  2.0 $&   295  & 0.804\\
  2MASS~1507  &  L5   & 2$\tablenotemark{c}$ & $0.5  \pm  0.9 $ & -2.0 & $2.5 \pm 1.3$& 1415 & 0.950\\
  \enddata \tablenotetext{a}{flare star} \tablenotetext{b}{only order
    41, 44} \tablenotetext{c}{only orders 41, 42}
\end{deluxetable}

\begin{deluxetable}{rlccrccr}
  \tablecaption{\label{tab:constant} Stars with probably constant radial velocities} 
  \tablewidth{0pt}
  \tablecolumns{8}
  \tablehead{Object & Spectral & No. of & $\Delta v_{\textrm {star}}$ & \colhead{$\Delta v_{\textrm {tell}}$} & \colhead{$\Delta v_{\textrm {total}}$} & max. $\Delta t$ & \colhead{$p$}\\ 
    & Type & Spectra & [km/s] & [km/s] & [km/s] & [d] &}
  \startdata
  LHS~3494    &  M5.5 & 3 & $0.4  \pm  0.6 $    &   -0.2 & $0.6  \pm  1.1 $  &   709 & 0.421\\
  LP~759      &  M5.5 & 3 & $0.9  \pm  0.7 $    &    1.7 & $0.8  \pm  1.1 $  &  1018 & 0.517\\
  LHS~2876    &  M6   & 3 & $0.3  \pm  0.5 $    &        & $0.3  \pm  1.4 $  &  1846 & 0.165\\
  LHS~292     &  M6   & 3 & $0.4  \pm  0.4 $    &        & $0.4  \pm  1.4 $  &  3506 & 0.224\\
  CTI~0004    &  M6   & 2 & $0.0  \pm  0.2 $    &    0.3 & $0.3  \pm  0.9 $  &   738 & 0.255\\
  CTI~0042    &  M6   & 2 & $1.8  \pm  4.0 $    &    2.4 & $0.6  \pm  4.1 $  &   739 & 0.116\\
  LP~731-47   &  M6.5 & 2 & $1.0  \pm  0.2 $    &    0.1 & $0.9  \pm  0.9 $  &  1459 & 0.671\\
  CTI~1539    &  M6.5 & 3 & $0.1  \pm  0.8 $    &        & $0.1  \pm  1.6 $  &   388 & 0.051\\
  LHS~523     &  M6.5 & 2 & $0.2  \pm  0.2 $    &        & $0.2  \pm  1.4 $  &  2903 & 0.117\\
  RX~2337     &  M7   & 2 & $0.5  \pm  0.3 $    &   -0.3 & $0.8  \pm  1.0 $  &   710 & 0.601\\  
  RX~0019     &  M7   & 2 & $0.5  \pm  0.9 $    &    0.6 & $0.1  \pm  1.3 $  &   709 & 0.063\\
  DENIS~2049  &  M7   & 3 & $1.6  \pm  0.3 $    &    0.9 & $0.7  \pm  1.0 $  &   337 & 0.539\\
  DENIS~2202  &  M7   & 3 & $0.6  \pm  0.6 $    &    0.2 & $0.4  \pm  1.1 $  &   337 & 0.288\\
  DENIS~2107  &  M7   & 3 & $1.5  \pm  0.1 $    &    2.1 & $0.6  \pm  0.9 $  &   336 & 0.492\\
  DENIS~2333  &  M7   & 2 & $0.1  \pm  0.1 $    &   -0.2 & $0.3  \pm  0.9 $  &   147 & 0.260\\
  LHS~1070    &  M7   & 3 & $0.8  \pm  0.1 $    &    0.5 & $0.3  \pm  0.9 $  &  3283 & 0.260\\
  LHS~3003    &  M7   & 2 & $0.2  \pm  0.3 $    &        & $0.2  \pm  1.4 $  &  3354 & 0.115\\
  CTI~1156    &  M7   & 3 & $0.5  \pm  0.3 $    &        & $0.5  \pm  1.4 $  &  1846 & 0.283\\
  VB~8        &  M7   & 4 & $1.2  \pm  0.2 $    &        & $1.2  \pm  1.4 $  &  2629 & 0.622\\
  2MASS~1254  &  M7.5 & 2 & $0.2  \pm  0.9 $    &    1.4 & $1.2  \pm  1.3 $  &  1847 & 0.654\\
  2MASS~1256  &  M7.5 & 2 & $0.9  \pm  1.5 $    &    0.3 & $0.6  \pm  1.8 $  &  1459 & 0.268\\
  LHS~2632    &  M7.5 & 3 & $0.9  \pm  0.1 $    &        & $0.9  \pm  1.4 $  &  2309 & 0.495\\
  LHS~2243    &  M8   & 2 & $0.1  \pm  0.1 $    &        & $0.1  \pm  1.4 $  &  2740 & 0.059\\
  LHS~2397A   &  M8   & 2 & $1.1  \pm  0.8 $    &        & $1.1  \pm  1.6 $  &  2630 & 0.517\\        
  RG~0050     &  M8   & 2 & $0.6  \pm  0.3 $    &        & $0.6  \pm  1.4 $  &  2535 & 0.336\\
  VB~10       &  M8   & 3 & $0.3  \pm  0.2 $    &        & $0.3  \pm  1.4 $  &  3759 & 0.174\\
  DENIS~2353  &  M8   & 2 & $2.1  \pm  0.4 $    &    2.0 & $0.1  \pm  1.0 $  &   145 & 0.081\\
  CTI~0126    &  M8.5 & 2 & $1.5  \pm  0.9 $    &        & $1.5  \pm  1.6 $  &  2535 & 0.645\\
  BRI~1222    &  M9   & 2 & $0.5  \pm  0.3 $    &        & $0.5  \pm  1.4 $  &  2630 & 0.283\\
  DENIS~1048  &  M9   & 3 & $1.1  \pm  0.3 $    &    0.4 & $0.7  \pm  1.0 $  &   721 & 0.539\\
  DENIS~2331  &  M9   & 2 & $0.2  \pm  0.1 $    &   -0.6 & $0.8  \pm  0.9 $  &   146 & 0.623\\
  LHS~2065    &  M9   & 3 & $1.7  \pm  0.5 $    &    1.2$\tablenotemark{d}$& $0.5  \pm  1.0 $ &  3760 & 0.373\\
  2MASS~2234  &  M9.5 & 2 & $0.3  \pm  0.6 $    &   -0.6 & $0.9  \pm  1.1 $  &   336 & 0.595\\
  2MASS~1439  &  L1   & 5 & $1.2  \pm  1.2 $    &    0.0 & $1.2  \pm  1.5 $  &  2096 & 0.576\\
  2MASS~1300  &  L1   & 2 & $0.3  \pm  0.6 $    &    0.4 & $0.1  \pm  1.1 $  &  1079 & 0.074\\
  2MASS~1656  &  L1   & 2 & $0.0  \pm  1.3 $    &   $-$1.2 & $1.2  \pm  1.6 $&   1077 & 0.552\\
  Kelu~1      &  L2   & 3$\tablenotemark{e}$ & $3.4  \pm  5.5 $    &    2.6 & $0.8  \pm  5.6 $&   358 & 0.114\\
  LHS~102B    &  L2   & 2 & $0.3  \pm  6.0 $    &    0.3 & $0.0  \pm  6.1 $  &  1135 & 0.000\\
  2MASS~1615  &  L3   & 2$\tablenotemark{f}$ & $2.4 $    &    1.2 & $1.2 \pm 1.6$ &  1415 & 0.552\\
  \enddata
  \tablenotetext{d}{from water bands}
  \tablenotetext{e}{only orders 41, 42, 45}
  \tablenotetext{f}{only order 44}
\end{deluxetable}

\section{Results}
\label{sec:results}

\begin{figure*}
  \mbox{ \includegraphics[width=0.3\hsize]{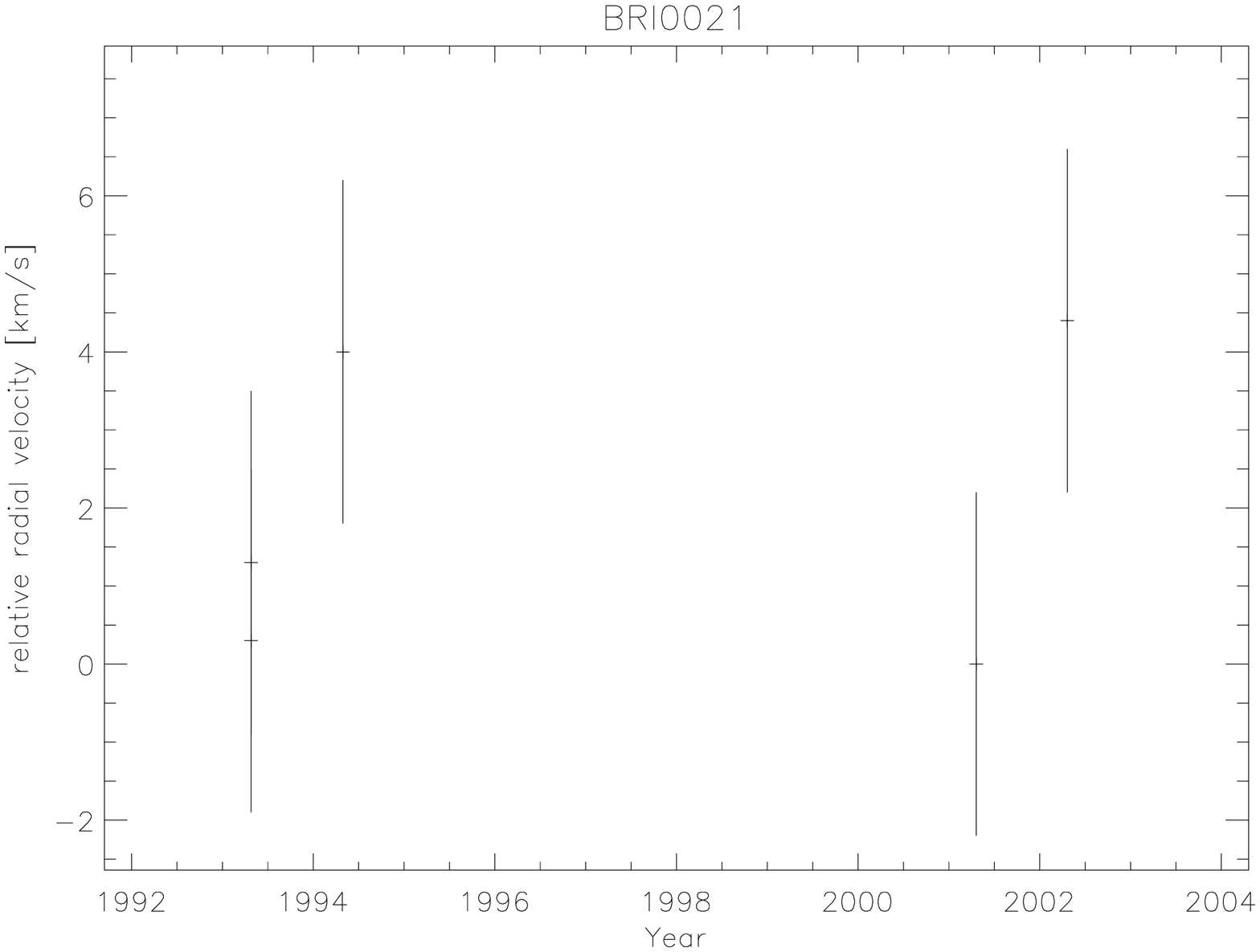}
    \includegraphics[width=0.3\hsize,bburx=668,bbury=488,bbllx=-20,bblly=-20]{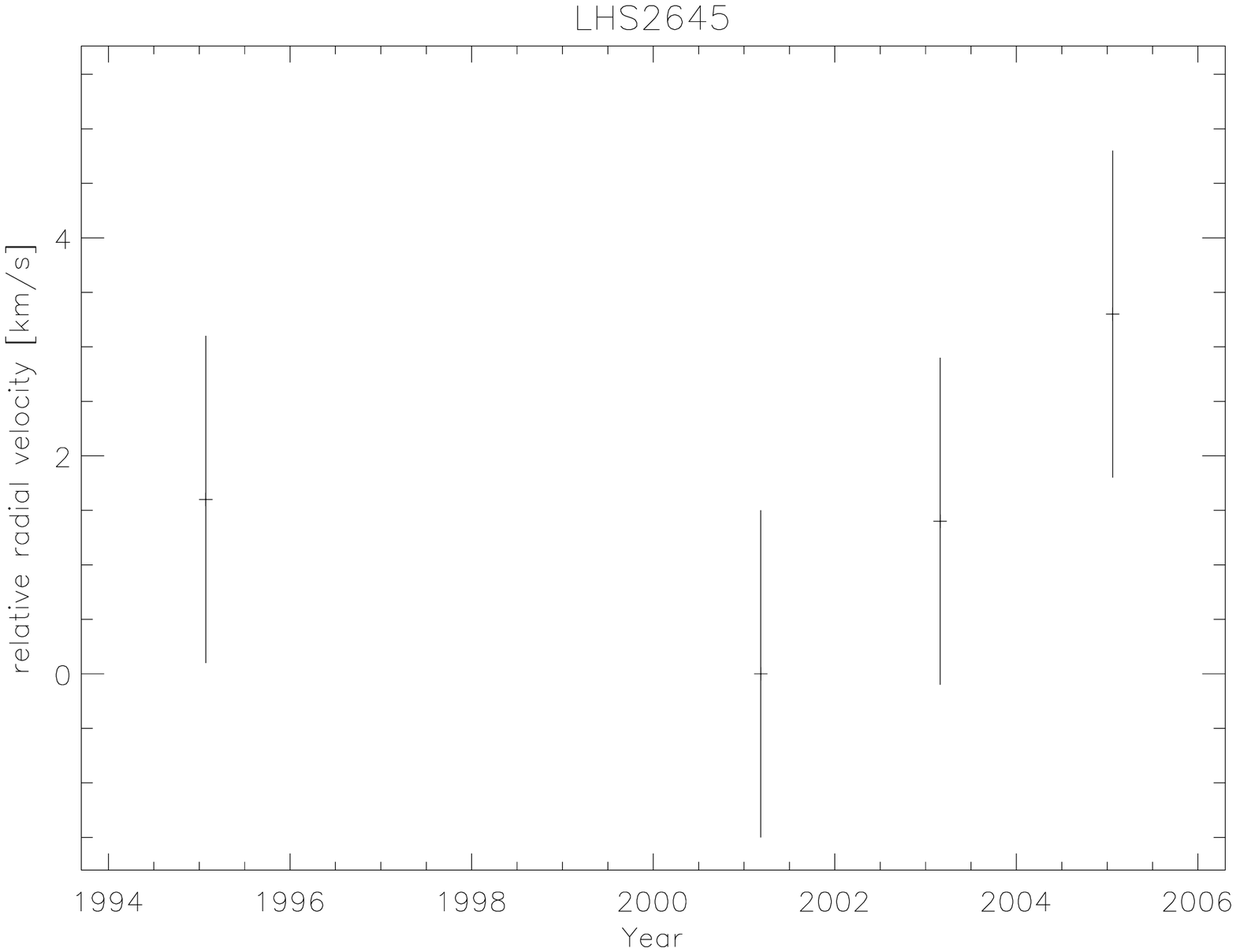}
    \includegraphics[width=0.3\hsize]{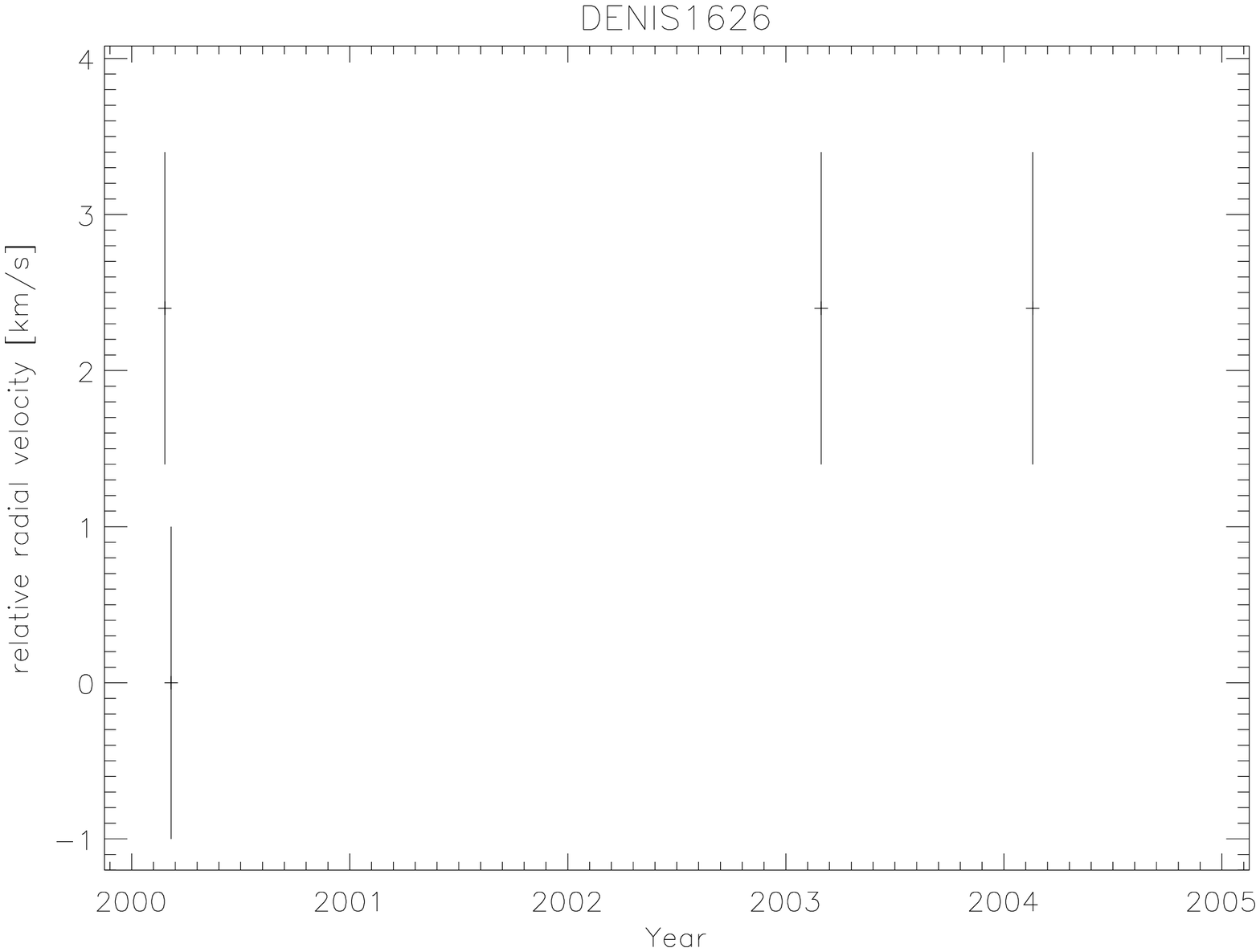} }
  \figcaption{\label{fig:radvels}Radial velocity variations of three
    of our targets. Left panel: BRI\,0021 (corrected for geometric
    shift); center panel: LHS\,2645 (no correction); right panel:
    DENIS\,1626 (corrected).}
\end{figure*}

The final results of our radial velocity study are in
Tables\,\ref{tab:variable} and \ref{tab:constant}. The object name and
spectral type are given in the first two columns, the number of
spectra taken is given in column three. Maximum radial velocity
variations between the spectra, $\Delta v_{\rm star}$, are calculated
by comparing the shift of the stellar spectrum to the shift of the
airglow lines. These values are given in column four. Uncertainties in
$\Delta v_{\rm star}$ are from the scatter of $\Delta v_{\rm star}$
calculated from different orders. We employed the four orders listed
in Table\,\ref{tab:orders} if not stated otherwise. To show how much
of the radial velocity shift is inferred from imperfect centering of
the star in the slit, we also give in column five the radial velocity
difference between the telluric bands in the different spectra.  If
available, the telluric shift was calculated from the oxygen A-band.
If no value is given, the A-band is not contained in our spectra and
we could not calculate a value (assuming zero in that case).  In one
case, the telluric reference was derived from water lines at $\lambda
= 7260$\,\AA. Column six, $\Delta v_{\rm total}$, contains the
difference between the velocity shift measured in the star and in the
telluric lines, this is our final radial velocity shift, with its
final error also listed. As noted above, an error of 0.9\,km\,s$^{-1}$
in the telluric position was added in quadrature to the error from
scatter in individual orders to produce the final error. The time
difference between our earliest and latest exposure for each star is
given in column seven, and the probability $p$ as discussed in the
previous section in column eight. In Fig.\,\ref{fig:radvels}, we show
the variation of radial velocity in three of our objects as an
example.

\subsection{Radial velocity variations of very low mass binaries}

In the previous section we have derived a binary frequency of
$11^{+0.07}_{-0.04}$\% for our observed sample. Now we
investigate the range of periods and semimajor axes our
measurements are sensitive to.  We can then compare our result to
predictions from star-formation scenarios and combine it with results
from different parameter spaces.

\begin{figure}
  \plottwo{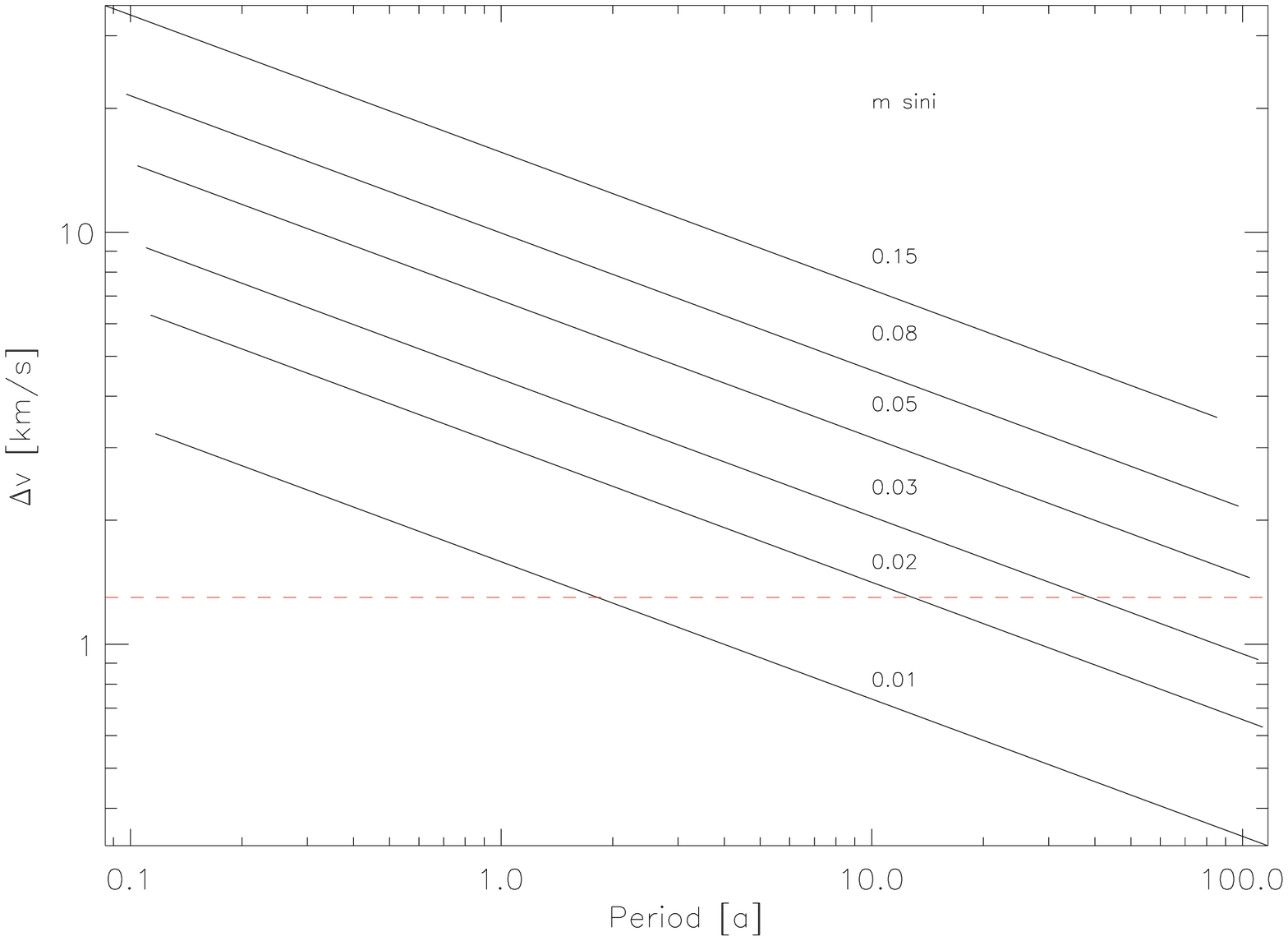}{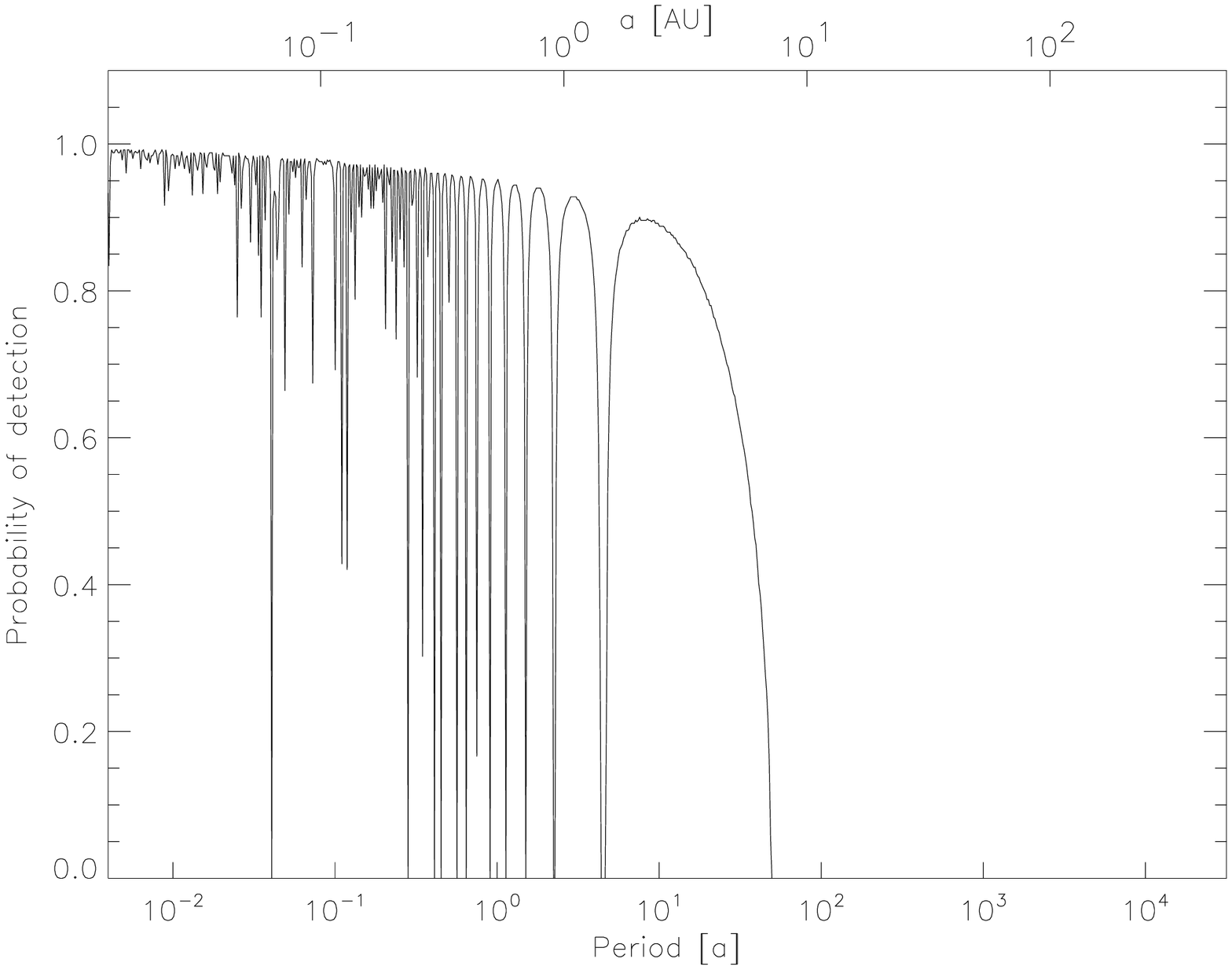} \figcaption{\label{fig:vrad_P}Left panel:
    Projected ($i=52^{\circ}$) variability of $v_{\rm rad}$ of a $M =
    0.15$\,M$_{\odot}$ primary ($\sim$M7) inferred by companions of $M
    = [0.15, 0.08, 0.05, 0.03, 0.02, 0.01]$\,M$_{\odot}$ as a function
    of orbital period. The dashed line indicates our average detection
    limit $\Delta v \approx 1.3$\,km\,s$^{-1}$. Right panel: Detection
    probability for a binary with $M_{\rm tot}=0.15$\,M$_{\odot}$,
    $q=M_2/M_1=0.5$, $i=52^{\circ}$ and $\Delta v =
    1.3$\,km\,s$^{-1}$.}
\end{figure}

Our sample consists of objects of spectral types between M5 and L5,
the majority being late M-dwarfs. In the left panel of
Fig.\,\ref{fig:vrad_P}, we show the projected radial velocity
variation of a primary with a mass of $M = 0.15$\,M$_{\odot}$
(corresponding approximately to an M7 field dwarf) that is inferred from
orbital motion of companions with masses $M =$ $[0.15, 0.08, 0.05, 0.03,
0.02, 0.01]$\,M$_{\odot}$, as indicated in the figure.  There is a
horizontal dashed line at $\Delta v = 1.3$\,km\,s$^{-1}$, our
approximate detection threshold. The radial velocity variations are
calculated for a system seen with an inclination of $i = 52\degr$.
The left panel of Fig.\,\ref{fig:vrad_P} shows that
binaries with an M7 primary can easily be detected even at small
mass-ratios.  Radial velocity variations become larger in later
objects and are even easier to detect. As opposed to planet hunting
surveys in G-type stars, where the accuracy has to be of the order of
a few meters per second, a precision of the order of a 
km\,s$^{-1}$ is sufficient in the case of very low mass primaries and
stellar companions.

The probability of detecting radial velocity variability in a binary
system depends in part on two factors: the accuracy of the measurement
and the time difference between two observations. The longer the time
separation, the larger the binary separation that can be found. In our
sample, the average longest time difference between two observations
of a target is 4.5\,yr (1645\,d). Assuming two observations separated
by 1645 days, we can calculate the probability that we can detect
radial velocity variability in a given binary system.  In the right
panel of Fig.\,\ref{fig:vrad_P}, we show this probability for the case
of a binary with a total mass of $M = 0.15$\,M$_{\odot}$, a mass ratio
of $q = M_2/M_1 = 0.5$, a detection threshold of $\Delta v =
1.3$\,km\,s$^{-1}$, and an angle between the orbital plane and the
line of sight of $i = 52\degr$. For a specific system, we can also
calculate the semimajor axis $a$ that corresponds to a given period,
$a$ is annotated at the upper border of that plot.

As can be seen in the right panel of Fig.\,\ref{fig:vrad_P}, the
distribution for the detection probability of a given binary system is
well above 90\% up to periods several times larger than the time
difference between observation epochs, $\Delta t$. It falls off
steeply towards longer periods. Several sharp gaps can be seen at
small periods; these occur when the two epochs of observation fall
within two parts of the orbit at which the primary has comparable
radial velocities.  The total fraction of binaries that remain
undetected due to such a combination of observing epochs and orbital
periods is smaller than 0.5\% of the total systems detectable; we will
neglect these gaps in the following. Third epoch observations or more
(between the two observations with the largest time difference) do not
have a significant effect on the probability distribution -- the main
effect is the filling in of the gaps. In the following, we will always
calculate the detection probabilities considering the first and last
epoch only (i.e., the largest time interval).

We do not know the parameters of the potential binaries we are
looking for, and in order to calculate the probability of finding any
binary, we have to account for the variety of free parameters
in such systems. \cite{MJ05} calculated the probability of detecting a
very low-mass binary using radial velocity variations of the primary
by performing a comprehensive Monte Carlo simulation. For the detection
probability of a binary system with given period $P$, they explicitly
took into account the distribution in primary mass $m$, mass ratio
$q$, eccentricity $e$, orbital phase $\Phi$, inclination $i$, and
longitude of periastron $\omega$. We analyze our observations
following the strategy of \cite{MJ05}, i.e.  calculating a detection
probability of binary systems given the accuracy of the measurement
and the time between observation epochs, $\Delta t$. For simplicity,
however, we do not perform a Monte Carlo simulation as
comprehensive as the one in \cite{MJ05}, but apply several
simplifications which are justified by their results:

\begin{itemize}
\item We use a fixed mass ratio of $q=0.5$ instead of 'flat' mass
  ratio distribution. \cite{MJ05} show the difference in detection
  probability between distributions is uniform in the range $q = 0.2 -
  1.0$ and uniform in the range $q = 0.7 - 1.0$. The difference turns
  out to be negligible. We arbitrarily selected $q=0.5$ for our
  calculations.
\item Binaries with periods less than 10\,d exhibit circular orbits
  \citep{MM05}. For systems with larger periods, \cite{MJ05} show
  detection probabilities for a distribution uniform in the range $e =
  0 - e_{\rm max}$ with $e_{\rm max} = 0.6$ and 0.9. No significant
  differences were found. We use $e = 0.0$ for our calculations.
\item We use a fixed primary mass of $M = 0.15$\,M$_{\odot}$. Although
  the detection probability would be somewhat different using the
  'correct' masses, no significant difference exceeding the
  uncertainty from our assumption of the mass ratio $q$ is expected in
  the overall detection probability.
\item The orbital phase $\Phi$ is held fixed in our calculations. The
  choice of $\Phi$ mainly determines the location of the gaps visible
  in the right panel of Fig.\,\ref{fig:vrad_P}. As mentioned above,
  the gaps cover an insignificant fraction of the probability
  distribution.
\item We use a fixed inclination of $i = 52\degr$, the mean
  inclination angle for arbitrarily oriented orbits.
\item Consideration of the longitude of periastron is unnecessary
  because of our choice of circular orbits.
\end{itemize}

\begin{figure}
\plotone{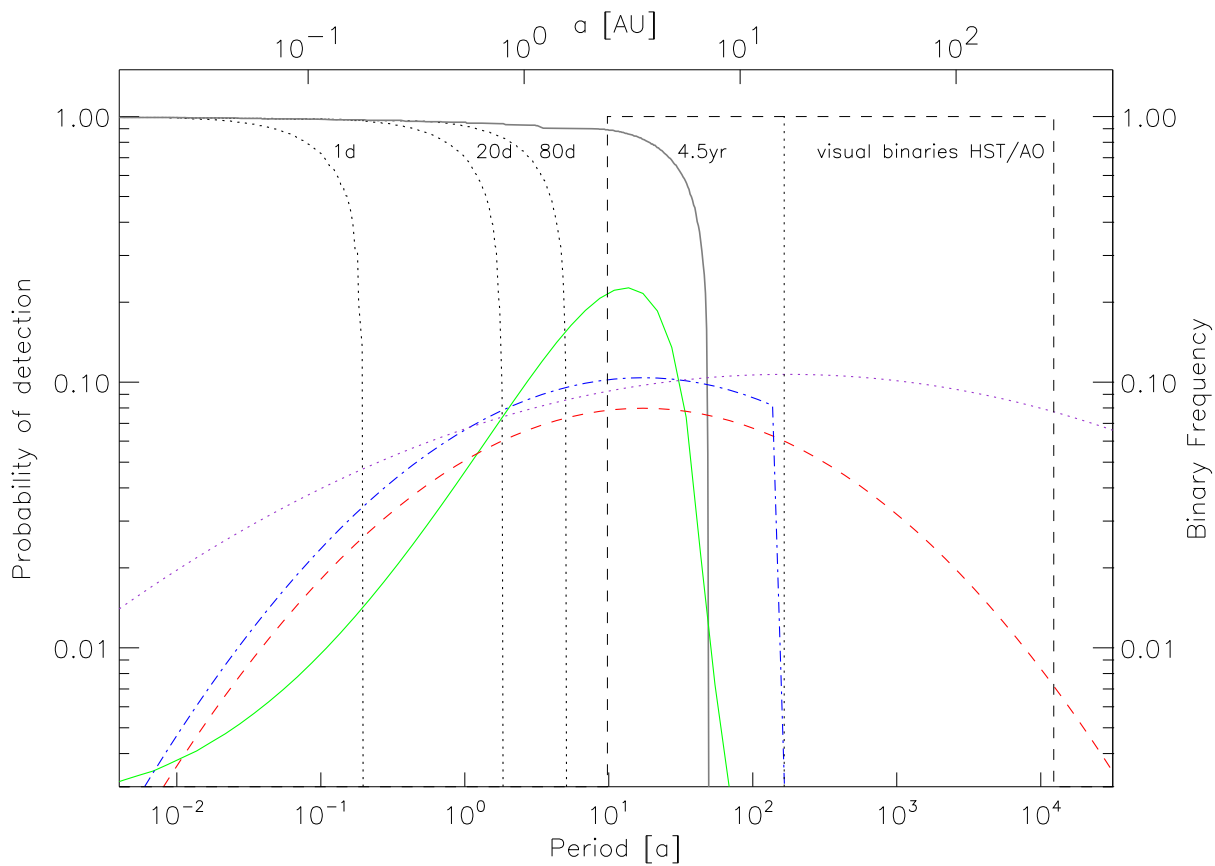}
\caption{\label{fig:Probabilities}Smoothed detection probabilities as 
  in Fig.\,\ref{fig:vrad_P} for four different $\Delta t$
  corresponding to recent surveys as explained in the text. The dashed
  box indicates the region sampled by imaging surveys (set to unit
  probablility in the region of sensitivity); the dotted vertical line
  is at 15 AU, beyond which no binaries have been found. Four possible
  binary distributions are overplotted in different colors -- see
  text.  }
\end{figure}

Under these assumption we can calculate the probability of detecting a
binary system for the data sets available for each of our targets. In
Fig.\,\ref{fig:Probabilities}, we show four probability distributions
of detecting a binary system under the assumptions explained above and
with $\Delta t$ = 1, 20, 80, and 1650\,d. The detection probabilities for
the four time differences correspond to sensitivities of the radial
velocity studies carried out by \citet[][$\Delta t = 1$\,d]{Ken05},
\citet[][$\Delta t = 20-80$\,d]{GW03, J05}, and this work with a mean
time difference of $\Delta t = 1650$\,d (4.5\,a). The corresponding
separations to which the different surveys are sensitive can be read
from the plotted curves. For a total mass of $M = 0.15$\,M$_{\odot}$
they are $a \la 0.2$\,AU, $a \la 1$\,AU, and $a \la 6$\,AU for $\Delta
t$ = 1\,d, 20--80\,d, and 1650\,d, respectively. We also plot
schematically the sensitivity of imaging surveys carried out with the
HST \citep{Close03} and AO \citep{Bouy03}; their detection probability
is of course also smaller than 1.0, which is neglected in the figure.
In the same plot, we also show four binary distributions that we will
test in the next chapter.

\clearpage

\clearpage

While \cite{Ken05, GW03} and \cite{J05} were sensitive only to periods
significantly smaller than the regime of the imaging surveys, the
survey presented here for the first time closes the gap between small
period binaries detected by radial velocity variations and resolved
large period binaries. Thus, measurements now cover the whole range of
binary separations and we are able to measure the total binary
frequency in very low mass objects. \cite{Close03} reports a binary
fraction of $15 \pm 7$\% for wider low mass binaries; their survey is
sensitive to separations between 2.6 and 300\,AU. Combined with our
result of a binary fraction of $11^{+0.07}_{-0.04}$\% for separations
smaller than $\la~6$\,AU, this sets an upper limit to the total binary
fraction in low mass objects to $26 \pm 10$\%. This is an upper limit
because the fraction of binaries in the region 2.6--6\,AU has been
counted twice. Our measurements do not provide individual separations
for our binary detections, and we don't know the fraction of
binaries in this range of separations.

\clearpage

\clearpage

\subsection{The separation distribution of very low mass binaries}

We can now compare a given binary fraction to the measurements by
multiplying the binary distribution with the probability of detection
for our sample. By using different surveys which are sensitive to
different separation ranges, we can also probe the binary distribution
as a function of separation. In the following, we test the four binary
distributions that are plotted in Fig.\,\ref{fig:Probabilities}:
\begin{enumerate}
\item The binary distribution found by \cite{DM91} for F- and G-dwarfs
  (the DM91 distribution, dotted purple line).
\item A compressed DM91 distribution (dashed red line). Compression of
  such a distribution is expected from simple scaling of the orbital
  parameters with mass. From the appendix in \cite{Fish04} one finds
  that in the log-normal DM91 distribution, both the dispersion and
  the peak scale proportionally to the system mass.
\item The compressed DM91 distribution as above, but truncated at
  15\,AU (dash-dotted blue line). This truncation is suggested by the
  observed lack of low mass binaries with separations larger than
  15\,AU.
\item A separation distribution following the results of
  \citet[][solid green line]{Umbreit05}. This distribution is a
  consequence of the decay of accreting low mass triple systems and
  can be interpreted as a prediction of the ejection scenario
  \citep{RC01}.  However, since a number of assumptions had to be made
  during the calculation of this distribution, and since the results
  depend on a limited sample of model calculations, the detailed shape
  of this distribution is rather unconstrained, especially at very low
  separations.
\end{enumerate}

In Fig.\,\ref{fig:Probabilities}, the distributions have been
normalized so that the total binary fraction of the DM91 distribution
is 62\% \citep{DM91}. The other three distributions plotted have a
total binary fraction of 26\% according to the result combined from
this survey and from \cite{Close03}.

We calculate the fraction of binaries expected from each of the four
distributions given the detection probabilities of each survey and
compare them to the number of binaries observed in the surveys. We
only use the radial velocity surveys of \cite{GW03} and \cite{J05},
the results from imaging surveys \citep{Close03, Bouy03}, and this
work.  For our survey, we individually calculate the probabilities for
the 53 observations and combine the results at the end, for the other
surveys we use the probability distribution as plotted in
Fig.\,\ref{fig:Probabilities}.

From the work of \cite{J05}, we adopt two binaries out of a sample of 
eleven. \citet{MJ05} adopt only one out of ten by restricting the sample to
very low masses. Our sample, and the sample of \cite{GW03}, includes
objects more massive than 0.1\,M$_{\odot}$, so we include the binary 
found at $M \approx 0.16$\,M$_{\odot}$. From the work of
\cite{GW03} we adopt only the object that shows radial velocity
variability out of a sample of 24 objects \citep[unlike][we do not 
modify the uncertainties reported in that paper]{MJ05}. Conservatively, 
we do not count the two spectroscopic binaries (SB2) found by \cite{GW03},
since our detection probability (to which we want to compare theirs)
only includes radial velocity variations rather than double lines. Admittedly,
however, if we had found any spectroscopic binaries we certainly would have
pursued them, and they would have been counted in our survey. 
Including the two SB2 would raise their binary fraction above the other
results (to 15\%), higher than any model tested at those small separations
(though not by a statistically significant amount). 

We do not use the results from \cite{Ken05}.  Due to their small time
separation, this survey is only sensitive to a very limited range in
separation (note the logarithmic scale in
Fig.\,\ref{fig:Probabilities}). Although the differences between the
binary distributions are more than a factor of two, they are of the
order of only 1\% in that region, which is greatly exceeded by the
statistical uncertainty of the survey. Furthermore, the accuracy of
their measurements is only 5\,km\,s$^{-1}$ and so not comparable to the
surveys of \cite{GW03}, \cite{J05} and this work. 

In Table\,\ref{tab:Models}, we present the fractions of binaries one
expects for each survey for different underlying binary distributions.
In the second row, we give the observational results to which the
expected numbers have to be compared. Two sets of normalizations
have been tested. In the first set the DM91 is normalized to its value
from F- and G-dwarfs (62\%), the other models are normalized to have a
15\% binary fraction at separations 2.6--15\,AU as observed in the
imaging surveys. The second set is normalized to a total binary
fraction of 26\%.

\begin{deluxetable}{cccccc}
  \tablecaption{\label{tab:Models} Comparison of the predictions of
    the four tested binary fractions to observations. Two sets of
    normalizations are shown; (a) normalization to 62\,\% total binary
    fraction for DM91 and 15\,\% HST/AO binary fraction, and (b)
    normalization to 26\,\% total binary fraction. }  
  \tablecolumns{6}
  \tablewidth{0pt}
  \tablehead{ & GW03/J05 & \colhead{this work} & \multicolumn{2}{c}{HST/AO imaging}  & total binary fraction\\
    {\footnotesize separation a} & {\footnotesize $<$ 1\,AU}&
    {\footnotesize $<$ 6\,AU}&{\footnotesize 2.6-15\,AU} &
    {\footnotesize 15--300\,AU} } \startdata
  Observations & $0.09^{+0.07}_{-0.05}$ & $0.11^{+0.07}_{-0.04}$ & $0.15 \pm 0.07$ & 0.00 & $<0.26 \pm 0.10$ \\
  \noalign{\medskip} \hline \noalign{\medskip}
  DM91            & 0.13 & 0.18 & 0.12  & 0.18 & {\bf 0.62}\\
  compressed DM91 & 0.12 & 0.19 & {\bf 0.15}  & 0.10 & 0.45 \\
  truncated comp. DM91 & 0.12 & 0.19 & {\bf 0.15} & 0.00 & 0.35\\
  U05             & 0.08 & 0.23 & {\bf 0.15}  & 0.00 & 0.35 \\
  \noalign{\medskip}
  DM91            & 0.05 & 0.08 & 0.05  & 0.08 & {\bf 0.26}\\
  compressed DM91 & 0.07 & 0.11 & 0.09  & 0.06 & {\bf 0.26} \\
  truncated comp. DM91 & 0.09 & 0.15 & 0.12 & 0.00 & {\bf 0.26}\\
  U05             & 0.06 & 0.17 & 0.11  & 0.00 & {\bf 0.26} \\
  \noalign{\medskip} \enddata
\end{deluxetable}

A comparison of the expected fraction of detections from binary
distributions to the observational results in Table\,\ref{tab:Models}
reveals that two distributions match the observations reasonably well:
the truncated compressed DM91-distribution and the distribution from
decaying triple systems in \cite{Umbreit05}. The DM91 distribution as
well as its compressed version yield too high a fraction of wide
binaries, as has been noted in several papers before. \cite{MJ05}
already found in their analysis of the results of \cite{GW03},
\cite{J05} and \cite{Ken05}, that a DM91 distribution truncated at
15\,AU is in agreement with the observations. We confirm this result
with our new survey covering a much larger part of the parameter
space. However, \cite{MJ05} also found that the distribution given in
\cite{Umbreit05} yields significantly too few binaries compared to the
observations. We find instead that the distribution from decaying
triple systems matches well the observed distribution of binaries.

The discrepancy between our result and the conclusions in \cite{MJ05}
may be due to the different construction of samples. We did not
include the sample of \cite{Ken05} for the reasons given above.
\cite{MJ05} construct one large sample out of the three surveys by
\cite{GW03}, \cite{J05} and \cite{Ken05}. For the distribution from
\cite{Umbreit05}, this results in a smaller average detection
probability since the binary fraction is very small at small
separation (and small periods). Furthermore, the uncertainties of the
Umbreit binary distribution are especially large at small separations.
We thus conclude that the current observations cannot rule out a binary
distribution as predicted from triple system decay in low mass
objects. The total binary fraction of $>$20\%, however, is a much
stronger argument against the ejection hypothesis as the dominant
formation scenario for low mass objects, as we discuss below.

\section{Discussion and Conclusions}
\label{sec:discuss}

Our primary purpose in this project was to test whether the inner
population of VLM binaries was hiding anything unexpected, given
that the outer population has a remarkable cutoff at around 15 AU.
We wondered whether the frequency of spectroscopic binaries was similar
to that of the resolved binaries or whether the ``missing'' binaries
can be found at small separations. Our sample was not chosen by 
reason of this search and so is rather ill-defined,
not meeting any criterion for completeness in either a magnitude or
volume-limited sense. It must therefore be viewed as a random sampling
of the true distribution, partially biased towards brighter objects. 
The well-known bias introduced by unresolved binaries is unlikely to
play a strong role here, since our objects span a large magnitude
range and were not selected primarily for their brightness (but instead
as they were discovered by a variety of surveys). Furthermore, there
are factors operating (like few epochs of observation per target and
non-optimal radial velocity precision) which may operate as bias in
the opposite direction. The best we can say is that our positive 
detections are likely to be meaningful.

Our basic result is that there may be nearly as many
($11^{+0.07}_{-0.04}$\%) spectroscopic binaries in the separation
range 0-6\,AU as resolved binaries ($15 \pm 7$\%) in the separation
range 2.5-300\,AU (though this distribution is in fact capped at
15\,AU). This leads us to an estimate of an upper limit of $26 \pm 10$\%
for the binary fraction of VLM objects (it is an upper limit because we
are not sure of the overlap fraction between the spectroscopic and
resolved populations). Indeed, more than half of the binaries listed
in \cite{Close03} are within the separation we are sensitive to,
although presumably we wouldn't detect all of them due to the
inclination distribution. Some indication that the overlap problem is
not extreme is given by the fact that the close binary frequency we
find here is not much greater than that found by \cite{MJ05}, who did
not sample out to the overlap distance.  One can reasonably conclude 
that the true binary fraction for VLM objects is between 20-25\%.

We have compared our inferred binary distribution with some plausible
possibilities (both empirical and theoretical). It is especially
germaine to compare the two binary separation probability
distributions that are compatible with all observations as shown in
Fig.\,\ref{fig:Probabilities}. The truncated compressed DM91 (tcDM91)
distribution and the distribution from \cite{Umbreit05} differ mostly
in the region outward of 5\,AU, where the tcDM91 is significantly
larger. The differences at smaller separations, where the Umbreit
distribution is smaller in the inner region and larger beyond 1\,AU
than the tcDM91 distribution are not significant in light of the large
uncertainties of the Umbreit distribution. The cutoff at large
separations in the tcDM91 distribution has been imposed for empirical
reasons, while in the theoretical Umbreit distribution the steep edge
is expected to be a strong function of total system mass. It would
thus not be surprising if the semiempirical tcDM91 and the theoretical
Umbreit distribution are two extreme forms of the underlying actual
situation: a DM91 distribution whose separation scale is decreased
according to mass \citep[see the Appendix]{Fish04} but also truncated
due to subsequent gravitational stripping of low-mass binary systems.

The more serious strike against the ejection hypothesis as a dominant
formation mode for VLM objects is the binary frequency itself. This
frequency is known to decrease as a function of stellar mass anyway.
It is over 70\% for OB stars, around 60\% for GK stars, and about 25\%
for M stars \citep{Delfosse04}. Thus, the fraction we find here is
just what would be expected if there were no change in the formation
mechanism between VLM objects and more massive stars. For field M
stars in the 0.25--0.5\,M$_\odot$ range, the frequency of wide binaries
in the 25--300\,AU range is about 12\% \footnote{This can be inferred 
  from Fig. 4 in \cite{Delfosse04}. We also checked by analyzing the 
  M binaries within 13pc tabulated by \cite{Poveda94} as
  a fraction of that total sample in the Gliese catalog.}, 
however, while that for the field VLM objects (0.05--0.2\,M$_\odot$) 
is nearly zero. \cite{Close03} have argued that the resolved VLMS 
binaries are more than an order of magnitude more tightly bound than the 
wide M binaries. Since their masses are not so different, one might resort 
to a formation-based explanation of this. On the other hand, the wide 
binary frequency drops by more than a factor of 3 in going from GK stars
($1.1 - 0.6$\,M$_\odot$) to field early-M stars ($0.5 - 0.25$\,M$_\odot$). 
The mass drops by another factor of two from the early-M to the late-M 
and L sample. If the wide binary frequency goes down by a similar factor 
again, small number statistics could lead to the observed results. 
The lack of wide binaries among field VLM objects, therefore, does not
necessarily provide strong support for a formation mechanism that is 
obviously different than for higher-mass stars.

The question of whether the gravitational destruction of wider VLM
binaries is intrinsic to their formation process or occurs after they
form is crucial to the evaluation of the ejection hypothesis as their
predominant mode of formation. Wide binaries are almost impossible to
preserve during ejection (one might very occasionally form them later
by capture). It is interesting, therefore, that a population of wide
binaries is beginning to turn up in examinations of very young
populations of VLM objects. \cite{Luh04} has reported one such system
in a star-forming region (projected separation 240\,AU) and
\cite{Billeres05} have found one in the field (projected separation
over 200\,AU). Very recently, \cite{Bouy05} have found at least one or
two wide binaries in the Upper Sco association. They claim a wide binary 
fraction of at least 5\%. If confirmed and extended via further
observations, this would strongly imply that the frequency of wide 
binaries in the field is an evolutionary rather than a formation issue.

In summary, our results for close binary VLM systems basically support
the hypothesis that they are an extension of the trends in higher-mass
stars. The number of spectroscopic binaries is what might be expected
from knowledge of the close resolved systems, and is also consistent
with the overall binary frequency of the next stellar mass range up.
We find no support in this for the proposition that VLM objects have a
substantially different formation mechanism, particularly one which 
depends on ejection from small-N clusters. It will be very difficult 
for that mechanism to account for the similar binary frequency in M 
stars. Alternatively, one might assert that stars form like brown
dwarfs; this is not entirely facetious in that it may be that small-N
clusters do play a role with stars. A common formation mechanism is 
bolstered by the large set of observations which suggest that the 
formation of individual VLMS objects displays all the characteristics 
of higher-mass T Tauri stars \citep{MJB05}. 

This survey has a number of inadequacies that mean our conclusions must
be viewed as tentative. The treatment of slit errors is open to criticism,
and our assessment of statistical significance can be argued with.
It would be good to conduct a more sensitive radial velocity variability 
survey on a better-constructed sample. 
The next set of investigations about the frequency of wider 
VLM binaries should concentrate on star-forming regions, to confirm 
whether their dearth in the field is initial or a result of evolution in 
the first few million years.

\goodbreak


\acknowledgments 

This work is based on observations obtained from the W.M. Keck
Observatory, which is operated as a scientific partnership among the
California Institute of Technology, the University of California and
the National Aeronautics and Space Administration. We would like to
acknowledge the great cultural significance of Mauna Kea for native
Hawaiians and express our gratitude for permission to observe from
atop this mountain. GB thanks the NSF for grant support through
AST00-98468. AR has received research funding from the European
Commission's Sixth Framework Programme as an Outgoing International
Fellow (MOIF-CT-2004-002544).






\clearpage

\appendix
\begin{deluxetable}{rrrc}
  \tablecaption{\label{tab:appendix}Differential radial velocities for
    all sets of observations. In each set, the bluemost observation is
    arbitrarily given a differential radial velocity of zero, and the
    other epochs have the difference in velocity relative to that one
    listed. The error given is the same as described in Section
    \ref{sec:analysis}, except that no telluric shift or error has been added.}  
  \tablecolumns{4} \tablewidth{0pt}
  \tablehead{ \colhead{Year} & \colhead{Month} & \colhead{Day} & \colhead{differential radial velocity}\\
      & & &[km\,s$^{-1}$]}
  \startdata

\multicolumn{4}{l}{2MASS0746}\\
2004 & 5 &11    &   0.0 $\pm$    1.8\\
2005 & 3 & 2    &   2.6 $\pm$    1.8\\[2mm]
\multicolumn{4}{l}{2MASS1242}\\
1998 & 5 &25     &  0.9 $\pm$    0.3\\
2002 & 5 &20     &  0.0 $\pm$    0.3\\[2mm]

\multicolumn{4}{l}{2MASS1254}\\
1998 & 5 &25     &  0.0 $\pm$    0.9\\
2003 & 6 &11     &  1.2 $\pm$    0.9\\[2mm]

\multicolumn{4}{l}{2MASS1256}\\
1998 & 5 &25     &  1.0 $\pm$    1.5\\
2002 & 5 &20     &  0.0 $\pm$    1.5\\[2mm]

\multicolumn{4}{l}{2MASS1300}\\
2000 & 6 &29     &  0.0 $\pm$    0.6\\
2003 & 6 &11     &  0.3 $\pm$    0.6\\[2mm]

\multicolumn{4}{l}{2MASS1439}\\
1999 & 6 &10     &  0.8 $\pm$    1.2\\
1999 & 6 &10     &  0.9 $\pm$    1.2\\
2000 & 6 &29     &  0.0 $\pm$    1.2\\
2004 & 5 &11     &  1.2 $\pm$    1.2\\
2005 & 3 & 2     &  0.8 $\pm$    1.2\\[2mm]

\multicolumn{4}{l}{2MASS1506}\\
2000 & 6 &29      & 4.6 $\pm$    1.3\\
2004 & 5 &11      & 0.0 $\pm$    1.3\\[2mm]

\multicolumn{4}{l}{2MASS1507}\\
2000 & 6 &29      & 0.0 $\pm$    1.5\\
2004 & 5 &11      & 2.5 $\pm$    1.3\\[2mm]

\multicolumn{4}{l}{2MASS1615}\\
2000 & 6 &29      & 1.2 $\pm$    0.0\\
2004 & 5 &11      & 0.0 $\pm$    0.0\\[2mm]

\multicolumn{4}{l}{DENIS1626}\\
2000 & 5 &31      & 2.4 $\pm$    0.3\\
2000 & 6 &29      & 0.0 $\pm$    0.3\\
2003 & 6 &11      & 2.4 $\pm$    0.3\\
2004 & 5 &11      & 2.4 $\pm$    0.3\\[2mm]

\multicolumn{4}{l}{2MASS1656}\\
2000 & 6 &29      & 1.2 $\pm$    1.3\\
2003 & 6 & 9      & 0.0 $\pm$    1.3\\[2mm]

\multicolumn{4}{l}{2MASS2234}\\
2003 & 6 &11     &  0.0 $\pm$    0.6\\
2004 & 5 &11     &  1.0 $\pm$    0.6\\[2mm]

\multicolumn{4}{l}{BRI~0021}\\
1993 & 11 &11    &   0.3 $\pm$    2.0\\
1993 & 11 &11    &   1.3 $\pm$    2.0\\
1994 & 11& 23    &   4.0 $\pm$    2.0\\
2001 & 10 &28    &   0.0 $\pm$    2.0\\
2002 & 10 &31    &   4.4 $\pm$    2.0\\[2mm]

\multicolumn{4}{l}{BRI~1222}\\
1995 & 3 &12     &  0.0 $\pm$    0.3\\
2002 & 5 &19     &  0.5 $\pm$    0.3\\[2mm]

\multicolumn{4}{l}{CTI~0004}\\
2001 & 10 &28    &   0.0 $\pm$    0.3\\
2003 & 11 & 3    &   0.1 $\pm$    0.3\\[2mm]

\multicolumn{4}{l}{CTI~0042}\\
2001 & 10 &27    &   0.6 $\pm$    4.0\\
2003 & 11 & 3    &   0.0 $\pm$    4.0\\[2mm]

\multicolumn{4}{l}{CTI~0126}\\
1994 & 11 &23    &   1.5 $\pm$    0.8\\
2001 & 10 &27    &   0.0 $\pm$    0.8\\[2mm]

\multicolumn{4}{l}{CTI~1156}\\
1998 & 5 &25     &  0.0 $\pm$    0.3\\
2003 & 6 & 9     &  0.4 $\pm$    0.3\\
2003 & 6 &10     &  0.0 $\pm$    0.3\\[2mm]

\multicolumn{4}{l}{CTI~1539}\\
2002 & 5 &19    &   0.0 $\pm$    0.8\\
2003 & 6 &10    &   0.1 $\pm$    0.8\\[2mm]

\multicolumn{4}{l}{DBD~0021}\\
1998 & 9 &20      & 3.1 $\pm$    0.7\\
2001 &10 &27     &  0.0 $\pm$    0.7\\[2mm]

\multicolumn{4}{l}{DENIS1048}\\
2000 & 5 &30    &   0.7 $\pm$    0.3\\
2000 & 5 &31    &   0.1 $\pm$    0.3\\
2002 & 5 &20    &   0.0 $\pm$    0.3\\[2mm]

\multicolumn{4}{l}{DENIS1148}\\
2000 & 5 &31    &   0.0 $\pm$    0.2\\
2003 & 6 &11     &  1.1 $\pm$    0.2\\[2mm]

\multicolumn{4}{l}{DENIS2049}\\
2003 & 6 &10    &   0.0 $\pm$    0.3\\
2003 &11 & 4    &   0.7 $\pm$    0.3\\
2004 & 5 &11    &   0.4 $\pm$    0.3\\[2mm]

\multicolumn{4}{l}{DENIS2107}\\
2003 & 6 &11    &   0.0 $\pm$    0.1\\
2003 &11 & 4    &   0.6 $\pm$    0.1\\
2004 & 5 &11    &   0.2 $\pm$    0.1\\[2mm]

\multicolumn{4}{l}{DENIS2202}\\
2003 & 6 &10     &  0.3 $\pm$    0.5\\
2003 &11 & 3    &   0.4 $\pm$    0.5\\
2004 & 5 &11    &   0.0 $\pm$    0.5\\[2mm]

\multicolumn{4}{l}{DENIS2331}\\
2003 & 6 &11    &   0.9 $\pm$    0.1\\
2003 &11 & 4    &   0.0 $\pm$    0.1\\[2mm]

\multicolumn{4}{l}{DENIS2333}\\
2003 & 6 &10      & 0.0 $\pm$    0.1\\
2003 &11 & 4      & 0.3 $\pm$    0.1\\[2mm]

\multicolumn{4}{l}{DENIS2353}\\
2003 & 6 &11      & 0.0 $\pm$    0.4\\
2003 &11 & 3      & 0.1 $\pm$    0.4\\[2mm]

\multicolumn{4}{l}{Kelu~1}\\
1997 & 6 & 2     &  0.6 $\pm$    5.5\\
1998 & 5& 24      & 0.9 $\pm$    5.5\\
1998 & 5 &25     &  0.0 $\pm$    5.5\\[2mm]

\multicolumn{4}{l}{LHS102b}\\
1998 & 9 &20    &   0.1 $\pm$    5.9\\
2001 &10 &27    &   0.0 $\pm$    5.9\\[2mm]

\multicolumn{4}{l}{LHS1070}\\
1993 &11 &11      & 0.3 $\pm$    0.1\\
1993& 11 &11      & 0.3 $\pm$    0.1\\
2002 &10 &31      & 0.0 $\pm$    0.1\\[2mm]

\multicolumn{4}{l}{LHS2065}\\
1994 &11 &23      & 0.5 $\pm$    0.5\\
2002 & 5 &19      & 0.0 $\pm$    0.5\\
2005 & 3 & 2      & 0.4 $\pm$    0.5\\[2mm]

\multicolumn{4}{l}{LHS2243}\\
1994 &11 &23      & 0.0 $\pm$    0.1\\
2002 & 5 &19      & 0.1 $\pm$    0.1\\[2mm]

\multicolumn{4}{l}{LHS2397a}\\
1995 & 3 &12      & 0.0 $\pm$    0.8\\
2002 & 5 &19      & 1.1 $\pm$    0.8\\[2mm]

\multicolumn{4}{l}{LHS248}\\
1993 &11 &11      & 0.9 $\pm$    1.3\\
1997 &12 & 7      & 3.0 $\pm$    1.3\\
2005 & 3 & 2      & 0.0 $\pm$    1.3\\[2mm]

\multicolumn{4}{l}{LHS2632}\\
1995 & 3 &12      & 0.9 $\pm$    0.1\\
2001 & 7 & 3      & 0.0 $\pm$    0.1\\[2mm]

\multicolumn{4}{l}{LHS2645}\\
1995 & 3 &13      & 1.6 $\pm$    0.6\\
2001 & 7 & 3      & 0.0 $\pm$    0.6\\
2003 & 6 & 9      & 1.4 $\pm$    0.6\\
2005 & 3 & 2      & 3.3 $\pm$    0.6\\[2mm]

\multicolumn{4}{l}{LHS2876}\\
1998 & 5 &24      & 0.0 $\pm$    0.5\\
2002 & 5 &20      & 0.3 $\pm$    0.5\\
2003 & 6 & 9      & 0.0 $\pm$    0.5\\[2mm]

\multicolumn{4}{l}{LHS292}\\
1993 &11 &11      & 0.1 $\pm$    0.4\\
2002 & 5 &19      & 0.0 $\pm$    0.4\\
2003 & 6 &11      & 0.3 $\pm$    0.4\\[2mm]

\multicolumn{4}{l}{LHS2924}\\
1995 & 3 &12      & 1.4 $\pm$    0.6\\
2000 & 6 &29      & 2.2 $\pm$    0.6\\
2001 & 7 & 3      & 1.7 $\pm$    0.6\\
2001 & 7 & 3      & 0.0 $\pm$    0.6\\[2mm]

\multicolumn{4}{l}{LHS3003}\\
1995 & 3 &13      & 0.0 $\pm$    0.3\\
2004 & 5 &11      & 0.2 $\pm$    0.3\\[2mm]

\multicolumn{4}{l}{LHS3339}\\
2003 & 6 &11      & 0.0 $\pm$    0.2\\
2004 & 5 &11      & 1.8 $\pm$    0.2\\[2mm]

\multicolumn{4}{l}{LHS3494}\\
2001 & 7 & 3      & 0.0 $\pm$    0.5\\
2003 & 6 &10      & 0.6 $\pm$    0.5\\[2mm]

\multicolumn{4}{l}{LHS3495}\\
2001 & 7 & 3      & 0.0 $\pm$    0.1\\
2003 & 6 &10      & 2.0 $\pm$    0.1\\[2mm]

\multicolumn{4}{l}{LHS523}\\
1994 &11 &25      & 0.0 $\pm$    0.2\\
2002 &10 &31      & 0.2 $\pm$    0.2\\[2mm]

\multicolumn{4}{l}{LP731-47}\\
1998 & 5 &25      & 0.0 $\pm$    0.2\\
2002 & 5 &20      & 1.0 $\pm$    0.2\\[2mm]

\multicolumn{4}{l}{LP759}\\
1998 & 9 &20      & 0.9 $\pm$    0.7\\
2001 & 7 & 2      & 0.0 $\pm$    0.7\\
2001 & 7 & 2      & 0.5 $\pm$    0.7\\[2mm]

\multicolumn{4}{l}{RG0050}\\
1994 &11 &23      & 0.0 $\pm$    0.3\\
2001 &10 &27      & 0.7 $\pm$    0.3\\[2mm]

\multicolumn{4}{l}{RX0019}\\
2001 & 7 & 2      & 0.0 $\pm$    0.8\\
2003 & 6 & 9      & 0.0 $\pm$    0.8\\[2mm]

\multicolumn{4}{l}{RX2337}\\
2001 & 7 & 2      & 0.8 $\pm$    0.3\\
2003 & 6 &10      & 0.0 $\pm$    0.3\\[2mm]

\multicolumn{4}{l}{VB10}\\
1994 &11 &25      & 0.3 $\pm$    0.2\\
1995 & 3 &12      & 0.0 $\pm$    0.2\\[2mm]

\multicolumn{4}{l}{VB8}\\
1995 & 3 &13      & 0.6 $\pm$    0.2\\
2001 & 7 & 3      & 0.0 $\pm$    0.2\\
2001 & 7 & 3      & 0.7 $\pm$    0.2\\
2002 & 5 &19      & 1.1 $\pm$    0.2\\[2mm]

\multicolumn{4}{l}{YZ~Cmi}\\
1994 &11 &25      & 0.0 $\pm$    0.1\\
2002 &10 &29      & 1.6 $\pm$    0.1\\[2mm]
\enddata
\end{deluxetable}

\clearpage


\end{document}